%% file: hats5.tex
\documentclass[apjl]{emulateapj}
\usepackage{comment}
\usepackage{ifthen}
\usepackage{lineno}

\usepackage{amsmath}

\input{newcommand_gen.tex}

\input{newcommand_spec.tex}

\newcommand{\hatcurisoshort}{YY}
\newcommand{\hatcurisocite}{yi:2001}
\newcommand{\hatcurlumind}{\arstar}
\newcommand{\hatcurjhkfilset}{ESO}

\newboolean{emulateapj}
\setboolean{emulateapj}{true}

\newboolean{rvtablelong}
\setboolean{rvtablelong}{true}

\newboolean{astroph}
\setboolean{astroph}{true}

\shortauthors{Zhou et al.}
\shorttitle{\hatcur\lowercase{b}}
\ifthenelse{\boolean{emulateapj}}{
    
}{
    
}
\ifthenelse{\boolean{emulateapj}}{
    \newcommand{\titlestar}{$\star$}
}{
    \newcommand{\titlestar}{\star}
}
\ifthenelse{\boolean{emulateapj}}{
    \newcommand{\titlestarstar}{$\star\star$}
}{
    \newcommand{\titlestarstar}{\star\star}
}

\begin{document}


\title{\hatcur\lowercase{b}: A Transiting Hot-Saturn from the HATSouth Survey  \altaffilmark{$\dagger$}}

\author{
G.~Zhou\altaffilmark{1},
D.~Bayliss\altaffilmark{1},
K.~Penev\altaffilmark{2,3},
G.~\'A.~Bakos\altaffilmark{2,3,\titlestar,\titlestarstar},
J.~D.~Hartman\altaffilmark{2,3},
A.~Jord\'an\altaffilmark{4},
L.~Mancini\altaffilmark{5},
M.~Mohler\altaffilmark{5},
Z.~Csubry\altaffilmark{2,3},
S.~Ciceri\altaffilmark{5},  
R.~Brahm\altaffilmark{4},  
M.~Rabus\altaffilmark{4},   
L.~Buchhave\altaffilmark{6},
T.~Henning\altaffilmark{5}, 
V.~Suc\altaffilmark{4},
N.~Espinoza\altaffilmark{4},
B.~B\'eky\altaffilmark{3},  
R.~W.~Noyes\altaffilmark{3},
B.~Schmidt\altaffilmark{1}, 
R.~P.~Butler\altaffilmark{7}, 
S.~Shectman\altaffilmark{8},
I.~Thompson\altaffilmark{8},
J.~Crane\altaffilmark{8},   
B.~Sato\altaffilmark{9}, 
B.~Cs\'ak\altaffilmark{5}, 
J.~L\'az\'ar\altaffilmark{10},
I.~Papp\altaffilmark{10},
P.~S\'ari\altaffilmark{10},
N.~Nikolov\altaffilmark{11,5}
}
\altaffiltext{1}{Research School of Astronomy and Astrophysics, Australian National University, Canberra, ACT 2611, Australia; email: george.zhou@anu.edu.au}

\altaffiltext{2}{Department of Astrophysical Sciences,
	Princeton University, NJ 08544, USA}

\altaffiltext{3}{Harvard-Smithsonian Center for Astrophysics,
	Cambridge, MA, USA}

\altaffiltext{$\star$}{Alfred P.~Sloan Research Fellow}

\altaffiltext{$\star\star$}{Packard Fellow}

\altaffiltext{4}{Departamento de Astronom\'ia y Astrof\'isica, Pontificia
	Universidad Cat\'olica de Chile, Av.\ Vicu\~na Mackenna 4860, 7820436 Macul,
	Santiago, Chile}

\altaffiltext{5}{Max Planck Institute for Astronomy, Heidelberg,
	Germany}

\altaffiltext{6}{Niels Bohr Institute, Copenhagen University, Denmark}

\altaffiltext{7}{Department of Terrestrial Magnetism, Carnegie Institution of Washington, 5241 Broad Branch Road NW, Washington, DC 20015-1305, USA}

\altaffiltext{8}{The Observatories of the Carnegie Institution of Washington, 813 Santa Barbara Street, Pasadena, CA 91101, USA}

\altaffiltext{9}{Department of Earth and Planetary Sciences, Tokyo Institute of Technology, 2-12-1 Ookayama, Meguro-ku, Tokyo 152-8551}

\altaffiltext{10}{Hungarian Astronomical Association, Budapest,
	Hungary}

\altaffiltext{11}{Astrophysics Group, School of Physics, University of Exeter, Stocker Road, Exeter EX4 4QL, UK}

\altaffiltext{$\dagger$}{
The HATSouth network is operated by a collaboration consisting of
Princeton University (PU), the Max Planck Institute f\"ur Astronomie
(MPIA), and the Australian National University (ANU).  The station at
Las Campanas Observatory (LCO) of the Carnegie Institute is operated
by PU in conjunction with collaborators at the Pontificia Universidad
Cat\'olica de Chile (PUC), the station at the High Energy Spectroscopic
Survey (HESS) site is operated in conjunction with MPIA, and the
station at Siding Spring Observatory (SSO) is operated jointly with
ANU.
}


\begin{abstract}

We report the discovery of \hatcurb, a transiting hot-Saturn orbiting a G type star, by the HATSouth survey. \hatcurb{} has a mass of $\mpl \approx \hatcurPPmshort$\,\mjup, radius of $\rpl \approx \hatcurPPrshort$\,\rjup, and transits its host star with a period of $P\approx\hatcurLCPshort$\,d.  The radius of \hatcurb{} is consistent with both theoretical and empirical models. The host star has a $V$ band magnitude of \hatcurCCapassmVshort, mass of \hatcurISOmshort\,\msun, and radius of \hatcurISOrshort\,\rsun. The relatively high scale height of \hatcurb{}, and the bright, photometrically quiet host star, make this planet a favourable target for future transmission spectroscopy follow-up observations. We reexamine the correlations in radius, equilibrium temperature, and metallicity of the close-in gas-giants, and find hot Jupiter-mass planets to exhibit the strongest dependence between radius and equilibrium temperature. We find no significant dependence in radius and metallicity for the close-in gas-giant population.

\setcounter{footnote}{0}
\end{abstract}



\section{Introduction}
\label{sec:introduction}

Transiting planets are the best characterised planets outside of our solar system. The transit geometry allow us to measure the mass, radius, and characterise the atmosphere \citep[e.g.][]{2002ApJ...568..377C,2005Natur.434..740D} and dynamics \citep[e.g.][]{2000A&amp;A...359L..13Q} of individual planets. As a result the of discoveries from wide-field ground and space-based photometric surveys \citep[e.g.][]{bakos:2004:hatnet,2006PASP..118.1407P,2010Sci...327..977B,bakos:2012:hatsouth}, statistical studies have revealed that close-in gas-giants are rare \citep[e.g.][]{2012ApJS..201...15H,2013ApJ...766...81F}, have relatively dark albedos \citep[e.g.][]{2011ApJ...729...54C}, and are found preferentially around metal-rich stars \citep[e.g.][]{2004A&amp;A...415.1153S,2012Natur.486..375B}.

Previous studies have also explored effect of irradiation and composition in inflating the radius of gas giants \citep[e.g.][]{2006A&amp;A...453L..21G,2011MNRAS.410.1631E,beky:2011,2012A&amp;A...540A..99E}. In particular, \citet{2011MNRAS.410.1631E,2012A&amp;A...540A..99E} found that the radius of Saturn-mass planets are more dependent on metallicity than Jupiter-mass planets, revealing a mass dependence to the inflation mechanisms. 

Intensive ground-based follow-up observations are extremely important for the characterisation of transiting gas-giants. Due to the mass degeneracy in the gas-giant regime, Saturns, Jupiters, and brown dwarfs cannot be distinguished from discovery transit photometry alone. The mass degeneracy is a major limitation against using the \emph{Kepler} candidate sample to study mass-dependent statistics of close-in gas-giant planets. The rarity of close-in gas giants, and the relative difficulty of characterising hot-Saturns compared to hot-Jupiters, leaves the hot-Saturn regime still poorly explored. Of the 299 confirmed transiting planets\footnote{exoplanets.org, 21 Dec 2013}, only 23 have masses in range of Saturn $(0.1 < M_p < 0.5 \,\mjup)$ and are found in close-in orbits $(P < 10\,\text{days})$. As a result, our statistical understanding of the hot-Saturn population is relatively less mature.

In this study, we report the discovery of the transiting hot-Saturn \hatcurb{} by the HATSouth survey. The HATSouth discovery, photometric and spectroscopic follow-up observations are detailed in \refsecl{obs}. Analyses of the results, including derivation of host star parameters, global modelling of the data, blend analyses, and constraints on the wavelength-radius relationship, are described in \refsecl{analysis}. In \refsecl{discussion}, we revisit some of the statistical trends for the close-in gas-giant population, and discuss \hatcurb{} in the context of the known hot-Saturns and hot-Jupiters. 

\section{Observations}
\label{sec:obs}

\subsection{Photometric detection}
\label{sec:detection}

The transit signal around \hatcur{} was first detected from photometric observations by the HATSouth survey \citep{bakos:2012:hatsouth}. HATSouth is a network of identical, fully-robotic telescopes located at three sites spread around the Southern Hemisphere, allowing continuous coverage of the surveyed fields. Altogether 8066 observations of \hatcur{} were obtained by the HATSouth units HS-1 in Chile, HS-3 in Namibia, and HS-5 in Australia from September 2009 to December 2010. Each unit consists of four 0.18\,m f/2.8 Takahasi astrographs and Apogee 4K$\times$4K U16M Alta CCD cameras. Each telescope has a field of view of $4^\circ \times 4^\circ$, with a pixel scale of $3.7\pxs$. The observations are performed with 4 minute exposures through the Sloan $r'$ filter. 

\begin{figure}[!ht]
\plotone{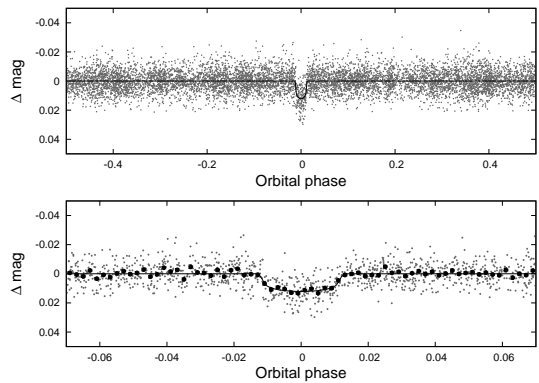}
\caption{
HATSouth \band{r'} discovery \lc, unbinned, and folded with a period of $P = \hatcurLCPprec$\,days, as per the analysis in \refsecl{analysis}. Solid line shows the best fit transit model. The lower panel shows the transit region of the \lc. Dark filled points represent the \lc{} binned at 0.002 in phase.
\label{fig:hatsouth}}
\end{figure}

Discussions of the HATSouth photometric reduction and candidate identification process can be found in detail in \citet{bakos:2012:hatsouth} and \citet{penev:2013:hats1}. Aperture photometry was performed and detrended using External Parameter Decorrelation \citep[EPD,][]{bakos:2007:hat2} and Trend Filtering Algorithm \citep[TFA,][]{kovacs:2005:TFA}. Transit signals were identified using the Box-fitting Least Squares analysis \citep[BLS,][]{kovacs:2002:BLS}. \reftabl{photobs} summarises the photometric observations for \hatcur{}. The HATSouth discovery \lc{} is plotted in \reffigl{hatsouth}. 

\ifthenelse{\boolean{emulateapj}}{
    \begin{deluxetable*}{llrrr}
}{
    \begin{deluxetable}{llrrr}
}
\tablewidth{0pc}
\tabletypesize{\scriptsize}
\tablecaption{
    Summary of photometric observations
    \label{tab:photobs}
}
\tablehead{
    \multicolumn{1}{c}{Facility}          &
    \multicolumn{1}{c}{Date(s)}             &
    \multicolumn{1}{c}{Number of Images}\tablenotemark{a}      &
    \multicolumn{1}{c}{Cadence (s)}\tablenotemark{b}         &
    \multicolumn{1}{c}{Filter}            \\
    &
    &
    &
    &
}
\startdata
HS-1 (Chile) & 2009 Nov--2010 Dec & 3953 & 290 & Sloan~$r$ \\
HS-3 (Namibia) & 2009 Sep--2010 Dec &  3241 & 288 & Sloan~$r$ \\
HS-5 (Australia) & 2010 Sep--2010 Dec & 900 & 287 & Sloan~$r$ \\
ESO/MPG 2.2\,m/GROND  & 2012 Oct 10 & 178 &  93 & Sloan~$g$ \\
ESO/MPG 2.2\,m/GROND  & 2012 Oct 10 & 107 &  93 & Sloan~$r$ \\
ESO/MPG 2.2\,m/GROND  & 2012 Oct 10 & 214 &  93 & Sloan~$i$ \\
ESO/MPG 2.2\,m/GROND  & 2012 Oct 10 & 211 &  93 & Sloan~$z$ \\
ESO/MPG 2.2\,m/GROND  & 2012 Dec 11 & 159 &  144 & Sloan~$g$ \\
ESO/MPG 2.2\,m/GROND  & 2012 Dec 11 & 160 &  144 & Sloan~$r$ \\
ESO/MPG 2.2\,m/GROND  & 2012 Dec 11 & 162 &  144 & Sloan~$i$ \\
ESO/MPG 2.2\,m/GROND  & 2012 Dec 11 & 162 &  144 & Sloan~$z$ \\
[-1.5ex]
\enddata 
\tablenotetext{a}{
  Outlying exposures have been discarded.
}
\tablenotetext{b}{
  Mode time difference between points in the \lc. Uniform sampling was not possible due to visibility, weather, pauses.
}
\ifthenelse{\boolean{emulateapj}}{
    \end{deluxetable*}
}{
    \end{deluxetable}
}

\subsection{Spectroscopy}
\label{sec:spec}

Spectroscopic confirmation of \hatcurb{} consisted of separate reconnaissance observations to exclude most stellar binary false-positive scenarios that can mimic the transit signal of an exoplanet. High resolution, high signal-to-noise measurements of the radial velocity (RV) variation for \hatcur{} were then obtained to confirm the planetary status of \hatcurb{}. The spectroscopic follow-up observations are presented in \reftabl{specobssummary}.

Low resolution reconnaissance observations were performed using the Wide Field Spectrograph \citep[WiFeS,][]{dopita:2007} on the ANU 2.3\,m telescope at Siding Spring Observatory, Australia. A flux calibrated spectrum was obtained at $R\equiv \lambda/\Delta \lambda =3000$ to provide an initial spectral classification of HATS-5 as an G-dwarf with $\teff = 5300\,\text{K}$, $\logg = 4.5$, and $\feh = 0$. These stellar parameters are later refined by higher resolution observations (Section~\ref{sec:analysis}). Multi-epoch observations at $R=7000$ confirmed the candidate did not exhibit $>1\,\kms$ RV variations. Such velocity variations are indicative of eclipsing stellar binaries, which have so far made up $\sim30\text{\%}$ of HATSouth candidates. Details of the WiFeS follow-up procedure and stellar binary identification process can be found in \citet{bayliss:2013:hats3} and \citet{2013arXiv1310.7591Z}. Candidates that pass the WiFeS vetting process are passed on to higher resolution observations.

\hatcur{} received nine high resolution ($R=60000$) reconnaissance RV observations with the CORALIE spectrograph on the Swiss Leonard Euler 1.2\,m telescope at La Silla Observatory, Chile, and fourteen $R=48000$ observations with the FEROS spectrograph on the ESO/MPG 2.2\,m telescope at La Silla. Detailed descriptions of the acquisition, reduction, and analyses of the CORALIE and FEROS observations can be found in the previous HATSouth discovery papers \citep[][]{penev:2013:hats1,mohlerfischer:2013:hats2}. Velocities from these observations allowed us to constrain the RV orbit semi-amplitude to be $<45\,\ms$.

\begin{figure} [ht]
\plotone{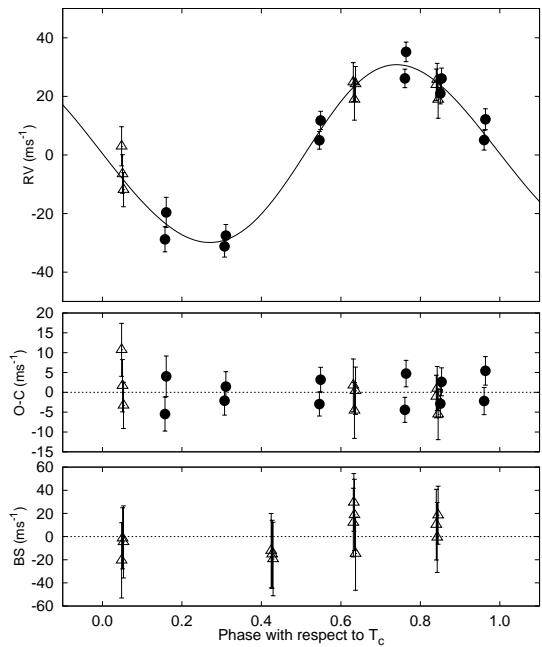}
\caption{
  {\em Top panel:} Phased radial velocities (RVs) from Magellan/PFS are plotted as dark filled circles, Subaru/HDS as open triangles. The best fit model is plotted by the solid line. The best fit absolute velocity offset from each instrument has been subtracted from the observations. {\em Middle panel:} Residuals of the RV measurements from the best fit model. The error bars have been inflated such that the $\chi^2$ per degree of freedom is unity for each instrument. {\em Bottom panel:} Bisector spans (BS) are plotted for velocities from Subaru/HDS. Note the different scales for each panel. 
\label{fig:rvbis}}
\end{figure}

The upper limit RV constraints from CORALIE, FEROS, and WiFeS indicated that \hatcurb{} is a low density gas-giant. High signal-to-noise, high resolution observations were required to determine the RV orbit of the system. Velocities of \hatcur{} were obtained Planet Finding Spectrograph (PFS) on the 6.5\,m Magellan Baade telescope at Las Campanas Observatory, Chile, and the High Dispersion Spectrograph (HDS) on the 8.2\,m Subaru telescope at Manua Kea Observatory, Hawaii. The PFS and HDS velocities and bisector spans are presented in \reftabl{rvs}, the RV orbit is plotted in \reffigl{rvbis}.

The Subaru/HDS \citep{2002PASJ...54..855N} observations were carried out on the nights of 19--22 Sep 2012 UT. Observations were made using an I$_{2}$ cell on four of the nights \citep{2002PASJ...54..865K}, and without the I$_{2}$ cell on one of the nights. We used the KV370 filter, the $0\farcs 6 \times 2\farcs 0$ slit, and the StdI2b setup, yielding spectra with a resolution of $R = 60000$ and wavelength coverage of 3500--6200\,\AA. On each night we obtained three consecutive observations yielding a total S/N per resolution element of $\sim 100$. The observations are split into three to reduce the impact of cosmic ray contamination and changes in the barycentric velocity correction over the course of an exposure. The I$_{2}$--free observations were used to create a template spectrum needed to measure precise relative RV values from the observations made with the I$_{2}$ cell. The individual spectra were reduced to RV measurements using the procedure of \citet{2002PASJ...54..873S,2012PASJ...64...97S}, which in turn is based on the method of \citet{1996PASP..108..500B}. Additionally we measured spectral line bisectors following \citet{bakos:2007:hat2} for each observation. The root-mean-squared (RMS) scatter of the HDS velocities from the best fit Keplerian curve is $4.8\,\ms$.

HATS-5 was also observed with the Carnegie Planet Finder Spectrograph \citep[PFS,][]{2010SPIE.7735E.170C} on Magellan~II at Las Campanas Observatory, Chile on the UT nights of 
2012 December 28-31,  2013 February 21, and 2013 February 4.  We obtained one iodine-free spectrum, and all other spectra were taken using the iodine cell and a slit-width of 0.5$\arcsec$.  To increase the signal-to-noise of each spectrum we read-out with 2$\times$2 binning and in slow readout mode.  Consecutive pairs of 20~min exposures were taken on each night.  The RV for each spectrum was determined using the spectral synthesis technique detailed in \citet{1996PASP..108..500B}. The (RMS) scatter of the PFS velocities from the best fit Keplerian curve is $3.8\,\ms$.

%

\ifthenelse{\boolean{emulateapj}}{
    \begin{deluxetable*}{llrrrrr}
}{
    \begin{deluxetable}{llrrr}
}
\tablewidth{0pc}
\tabletypesize{\scriptsize}
\tablecaption{
    Summary of spectroscopic observations\label{tab:specobssummary}
}
\tablehead{
    \multicolumn{1}{c}{Telescope/Instrument} &
    \multicolumn{1}{c}{Date Range}          &
    \multicolumn{1}{c}{Number of Observations} &
    \multicolumn{1}{c}{Resolution}          &
    \multicolumn{1}{c}{Observing Mode}          \\
}
\startdata
\sidehead{Reconnaissance}
ANU 2.3\,m/WiFeS & 2012 Aug 4 & 1 & 3000 & RECON Spec\tablenotemark{a}\\
ANU 2.3\,m/WiFeS & 2012 Aug 4--6 & 3 & 7000 & RECON RV\tablenotemark{b}\\
Euler 1.2\,m/Coralie & 2012 Aug 21--2013 Feb 27 & 9 & 60000 & ThAr/RECON RV \\
ESO/MPG 2.2\,m/FEROS & 2012 Nov 21--2013 Feb 27 & 14 & 48000 & ThAr/RECON RV \\
\sidehead{High resolution radial velocity}
Subaru 8.2\,m/HDS & 2012 Sep 20--22 & 9 & 60000 & I$_2$/RV\tablenotemark{c}\\
Magellan 6.5\,m/PFS & 2012 Dec 28--2013 Mar 4 & 12 & 76000 & I$_2$/RV \\
[-1.5ex]
\enddata 
\tablenotetext{a}{
  Reconnaissance observations used for initial spectral classifications
}
\tablenotetext{b}{
  Reconnaissance observations used to constrain the radial velocity variations
}
\tablenotetext{c}{
  High precision radial velocities to determine the spectroscopic orbit of the planet
}
\ifthenelse{\boolean{emulateapj}}{
    \end{deluxetable*}
}{
    \end{deluxetable}
}

\ifthenelse{\boolean{emulateapj}}{
    \begin{deluxetable*}{lrrrrrr}
}{
    \begin{deluxetable}{lrrrrrr}
}
\tablewidth{0pc}
\tablecaption{
    Relative radial velocities and bisector span measurements of
    \hatcur{}.
    \label{tab:rvs}
}
\tablehead{
    \colhead{BJD} & 
    \colhead{RV\tablenotemark{a}} & 
    \colhead{\ensuremath{\sigma_{\rm RV}}\tablenotemark{b}} & 
    \colhead{BS} & 
    \colhead{\ensuremath{\sigma_{\rm BS}}} & 
        \colhead{Phase} &
        \colhead{Instrument}\\
    \colhead{\hbox{(2\,400\,000$+$)}} & 
    \colhead{(\ms)} & 
    \colhead{(\ms)} &
    \colhead{(\ms)} &
    \colhead{} &
        \colhead{} &
        \colhead{}
}
\startdata
\input{data/rvtable.tex}
    [-1.5ex]
\enddata
\tablenotetext{a}{
  An instrumental offset in the velocities $(\gamma_{\rm rel})$ from each instrument was fitted for and subtracted in the analysis and the values presented in this table. Observations without an RV measurement are I$_2$-free template observations, for which only the bisector (BS) is measured. 
}
\tablenotetext{b}{
        Internal errors excluding the component of
        astrophysical/instrumental jitter considered in
        \refsecl{analysis}.
}
\tablenotetext{c}{
  HDS template observations made without the Iodine cell. We only measure the BS values for these observations.
}
\ifthenelse{\boolean{emulateapj}}{
    \end{deluxetable*}
}{
    \end{deluxetable}
}

\subsection{Photometric follow-up observations}
\label{sec:phot}

\begin{figure*}[!ht]
\plotone{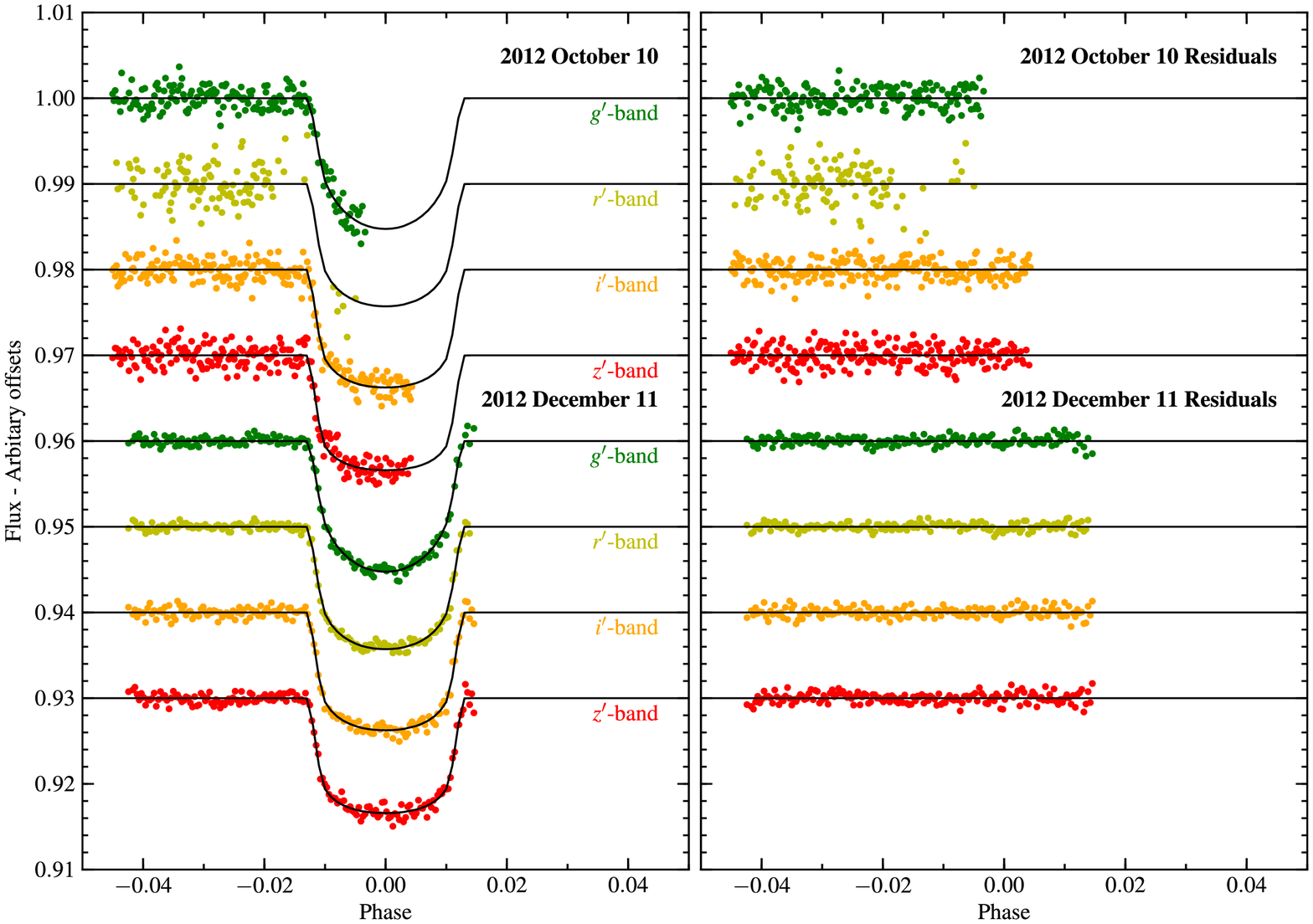}

\caption{
  {\em Left:} GROND follow-up transit \lcs{} in the $g'$-, $r'$-, $i'$- and \band{z'} are plotted. The \lcs{} have been treated with EPD simultaneous to the transit fitting (Section~\ref{sec:analysis}). The best fit model is plotted as a solid line for each observation. {\em Right:} Residuals for each transit observation is plotted.
\label{fig:lc}} \end{figure*}

High precision photometric follow-ups of a partial and a full transit of \hatcurb{} were performed on 2012 October 10 and 2012 December 11, respectively, using GROND on the ESO/MPG 2.2\,m telescope \citep{2008PASP..120..405G}. The GROND imager provides simultaneous photometric monitoring in four optical bands $(g',r',i',z')$ over a $5.4'\times5.4'$ field of view at $0.158\,\pxs$ sampling. Details of the GROND observation strategy, reduction, and photometry procedure can be found in \citet{penev:2013:hats1} and \citet{mohlerfischer:2013:hats2}. The GROND \lcs{} are presented in \reftabl{phfu} and plotted in \reffigl{lc}.

\ifthenelse{\boolean{emulateapj}}{
        \begin{deluxetable*}{lrrrrr} }{
        \begin{deluxetable}{lrrrrr} 
    }
        \tablewidth{0pc}
        \tablecaption{Differential photometry of
        \hatcur\label{tab:phfu}} \tablehead{ \colhead{BJD} &
        \colhead{Mag\tablenotemark{a}} &
        \colhead{\ensuremath{\sigma_{\rm Mag}}} &
        \colhead{Mag(orig)\tablenotemark{b}} & \colhead{Filter} &
        \colhead{Instrument} \\ \colhead{\hbox{~~~~(2\,400\,000$+$)~~~~}}
        & \colhead{} & \colhead{} & \colhead{} & \colhead{} &
        \colhead{} } \startdata \input{data/phfu_tab_short.tex}
        [-1.5ex]
\enddata \tablenotetext{a}{
  Magnitudes have the out-of-transit level subtracted. HATSouth magnitudes (HS) have been treated with EPD and TFA prior to the transit fitting. The detrending and potential blending may cause the HATSouth transit to be up to 8\% shallower than the true transit. Follow-up \lcs{} from GROND have been treated with EPD simultaneous to the transit fitting. 
}
\tablenotetext{b}{
  Pre-EPD magnitudes are presented for the follow-up \lcs.
}
\tablecomments{
        This table is available in a machine-readable form in the
        online journal.  A portion is shown here for guidance regarding
        its form and content.
} \ifthenelse{\boolean{emulateapj}}{ \end{deluxetable*} }{ \end{deluxetable} }

\section{Analysis}
\label{sec:analysis}

The stellar parameters for \hatcur{} are derived from the PFS iodine-free spectrum using the Stellar Parameter Classiﬁcation (SPC) process described in \citet{2012Natur.486..375B}. The derived values for effective temperature, surface gravity, metallicity, and projected rotational velocity are $\teff=5300\pm50\,\text{K}$, $\logg=\hatcurSMEilogg\,\text{cgs}$,  $\feh=\hatcurSMEizfeh\,\text{dex}$, and $\vsini=\hatcurSMEivsin\,\kms$, respectively. The surface gravity is later confirmed from transit light curve fitting as per \citet{sozzetti:2007}. The SPC derived stellar parameters agree with the classifications made by the reconnaissance spectroscopic observations to within $10\,\text{K}$ in \teff, 0.4 dex in \logg, and 0.2 dex in \feh. 

To derive the system parameters, we performed a global analysis of the HATSouth discovery \lcs{}, follow-up photometry from GROND, and RV orbit measurements from PFS and HDS. The best fit parameters and posteriors are determined using a Markov chain Monte Carlo analysis, the global analysis procedure is fully described in \citet{bakos:2010:hat11} and \citet{penev:2013:hats1}. Following \citet{sozzetti:2007}, we use the stellar density from the \lc{} in the global fit and the spectroscopic stellar parameters, \teff{} and \feh, to sample from the Yonsei-Yale theoretical isochrones \citep{yi:2001}, deriving the stellar mass and radius for \hatcur{}. The resulting \logg{} from the isochrone sampling matches the spectroscopic \logg{} from SPC. The full list of final spectroscopic and derived stellar properties are presented in \reftabl{stellar}, the fitted system parameters and derived planet properties in \reftabl{planetparam}.

To rule out the possibility that HATS-5 is a blended eclipsing stellar binary system, rather than a transiting planet system, we carried out a blend analysis following \citet{hartman:2011:hat3233}. Based on the light curves, spectroscopically determined atmospheric parameters, and absolute photometry, we are able to exclude scenarios involving a stellar binary blended with a third star (either physically associated, or not associated with the binary) with 7$\sigma$ confidence. In order to fit the light curves, the blend scenarios require a combination of stars with redder broad-band colours than are observed. Moreover, the best-fit blend model would produce RV variations of several km\,s$^{-1}$ and bisector variations of several hundred m\,s$^{-1}$, which are substantially greater than the observed variations.  We conclude that the observations of HATS-5 are best explained by a model consisting of a planet transiting a star.

To search for rotational modulations of the host star, we perform a Lomb-Scargle \citep{1976Ap&amp;SS..39..447L,1982ApJ...263..835S} analysis of the HATSouth discovery \lcs{}, with the transits masked. No statistically significant peaks were identified in the TFA \lcs. The expected rotation period from the spectroscopic \vsini{} measurement is 34 days, which is difficult to measure from ground-based photometry (most of the HATSouth photometric data for \hatcurb{} were gathered over $\sim3$ months). We find no emission features in the Calcium H and K lines in the iodine free HDS and PFS spectra, indicating minimal chromospheric activity. The slow rotation rate and the lack of chromospheric activity are both consistent with the isochrone age estimate for \hatcur{}.


\begin{figure}[!ht]
\plotone{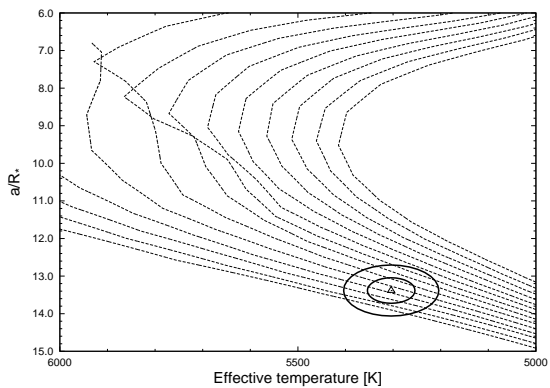}
\caption{
  Model isochrones from \cite{\hatcurisocite} for \hatcur{} are plotted. The isochrones for \feh = +\hatcurSMEizfehshort, ages of 0.2\,Gyr (lowest dashed line), and 1 to 13\,Gyr in 1\,Gyr increments are shown from left to right. The SPC values for $\teffstar$ and \arstar\ are marked by the open triangle, the $1\sigma$ and $2\sigma$ confidence ellipsoids are marked by solid lines.
\label{fig:iso}}
\end{figure}


\begin{deluxetable*}{lrl}
\tablewidth{0pc}
\tabletypesize{\scriptsize}
\tablecaption{
    Stellar parameters for \hatcur{}
    \label{tab:stellar}
}
\tablehead{
    \multicolumn{1}{c}{~~~~~~~~Parameter~~~~~~~~}   &
    \multicolumn{1}{c}{Value} &
    \multicolumn{1}{c}{Source}
}
\startdata
\noalign{\vskip -3pt}
\sidehead{Catalogue Information}
~~~~GSC \dotfill & 5897-00933 & \\
~~~~2MASS \dotfill& 04285348-2128548 & \\
~~~~RA (J2000) \dotfill & \hatcurCCra{} & 2MASS\\
~~~~DEC (J2000) \dotfill& \hatcurCCdec{} & 2MASS\\
\sidehead{Spectroscopic properties}
~~~~$\teffstar$ (K)\dotfill         &  \hatcurSMEteff & SPC\tablenotemark{a}\\
~~~~$\feh$\dotfill                  &  \hatcurSMEzfeh & SPC                 \\
~~~~$\vsini$ (\kms)\dotfill         &  \hatcurSMEvsin & SPC                 \\
\sidehead{Photometric properties}
~~~~$V$ (mag)\dotfill               &  \hatcurCCapassmV & APASS                \\
~~~~$B$ (mag)\dotfill               &  \hatcurCCapassmB & APASS                \\
~~~~$J$ (mag)\dotfill               &  \hatcurCCtwomassJmag & 2MASS           \\
~~~~$H$ (mag)\dotfill               &  \hatcurCCtwomassHmag & 2MASS           \\
~~~~$K_s$ (mag)\dotfill             &  \hatcurCCtwomassKmag & 2MASS           \\
\sidehead{Derived properties}
~~~~$\mstar$ ($\msun$)\dotfill      &  \hatcurISOmlong & \hatcurisoshort+\hatcurlumind+SPC \tablenotemark{b}\\
~~~~$\rstar$ ($\rsun$)\dotfill      &  \hatcurISOrlong & \hatcurisoshort+\hatcurlumind+SPC         \\
~~~~$\loggstar$ (cgs)\dotfill       &  \hatcurISOlogg & \hatcurisoshort+\hatcurlumind+SPC         \\
~~~~$\lstar$ ($\lsun$)\dotfill      &  \hatcurISOlum & \hatcurisoshort+\hatcurlumind+SPC         \\
~~~~$M_V$ (mag)\dotfill             &  \hatcurISOmv & \hatcurisoshort+\hatcurlumind+SPC         \\
~~~~$M_K$ (mag,\hatcurjhkfilset)\dotfill &  \hatcurISOMK & \hatcurisoshort+\hatcurlumind+SPC         \\
~~~~Age (Gyr)\dotfill               &  \hatcurISOage  & \hatcurisoshort+\hatcurlumind+SPC         \\
~~~~Distance (pc)\dotfill           &  \hatcurXdist   & \hatcurisoshort+\hatcurlumind+SPC\\
[-1.5ex]
\enddata
\tablenotetext{a}{
  SPC: The stellar parameters are derived from the PFS iodine-free spectrum using the Stellar Parameter Classification (SPC) pipeline \citep{2012Natur.486..375B}. These parameters also have small dependences on the global model fit and isochrone search iterations.
}
\tablenotetext{b}{
    \hatcurisoshort+\hatcurlumind+SPC: Based on the \hatcurisoshort\
    isochrones \citep{\hatcurisocite}, \hatcurlumind\ as a luminosity
    indicator, and the SPC results.
}
\end{deluxetable*}

\begin{deluxetable*}{lr}
\tabletypesize{\scriptsize}
\tablecaption{Orbital and planetary parameters\label{tab:planetparam}}
\tablehead{
        \multicolumn{1}{c}{~~~~~~~~~~~~~~~Parameter~~~~~~~~~~~~~~~} &
        \multicolumn{1}{c}{Value} \\
}
\startdata
\noalign{\vskip -3pt}
\sidehead{\Lc{} parameters}
~~~$P$ (days)             \dotfill    & $\hatcurLCP$ \\
~~~$T_c$ (${\rm BJD}$)    
      \tablenotemark{a}   \dotfill    & $\hatcurLCT$ \\
~~~$T_{14}$ (days)
      \tablenotemark{a}   \dotfill    & $\hatcurLCdur$ \\
~~~$T_{12} = T_{34}$ (days)
      \tablenotemark{a}   \dotfill    & $\hatcurLCingdur$ \\
~~~$\arstar$              \dotfill    & $\hatcurPPar$ \\
~~~$\zrstar$\tablenotemark{b}              \dotfill    & $\hatcurLCzeta$ \\
~~~$\rpl/\rstar$          \dotfill    & $\hatcurLCrprstar$ \\
~~~$b \equiv a \cos i/\rstar$
                          \dotfill    & $\hatcurLCimp$ \\
~~~$i$ (deg)              \dotfill    & $\hatcurPPi$ \\

\sidehead{Limb-darkening coefficients \tablenotemark{c}}
~~~$a_r$ (linear term)   \dotfill    & $\hatcurLBir$ \\
~~~$b_r$ (quadratic term) \dotfill    & $\hatcurLBiir$ \\
~~~$a_R$                 \dotfill    & $\hatcurLBiR$ \\
~~~$b_R$                 \dotfill    & $\hatcurLBiiR$ \\
~~~$a_i$                 \dotfill    & $\hatcurLBii$ \\
~~~$b_i$                  \dotfill    & $\hatcurLBiii$ \\

\sidehead{RV parameters}
~~~$K$ (\ms)              \dotfill    & $\hatcurRVK$ \\
~~~$\sqrt{e}\cos\omega$ 
                          \dotfill    & $\hatcurRVrk$ \\
~~~$\sqrt{e}\sin\omega$
                          \dotfill    & $\hatcurRVrh$ \\
~~~$e\cos\omega$ 
                          \dotfill    & $\hatcurRVk$ \\
~~~$e\sin\omega$
                          \dotfill    & $\hatcurRVh$ \\
~~~$e$                    \dotfill    & $\hatcurRVeccen$ \\
~~~$\omega$                    \dotfill    & $\hatcurRVomega$ \\
~~~PFS RV jitter (\ms)\tablenotemark{d}        
                          \dotfill    & $\hatcurRVjitterA$ \\
~~~HDS RV jitter (\ms)        
                          \dotfill    & $\hatcurRVjitterB$ \\

\sidehead{Planetary parameters}
~~~$\mpl$ ($\mjup$)       \dotfill    & $\hatcurPPmlong$ \\
~~~$\rpl$ ($\rjup$)       \dotfill    & $\hatcurPPrlong$ \\
~~~$C(\mpl,\rpl)$
    \tablenotemark{e}     \dotfill    & $\hatcurPPmrcorr$ \\
~~~$\rhopl$ (\gcmc)       \dotfill    & $\hatcurPPrho$ \\
~~~$\log g_p$ (cgs)       \dotfill    & $\hatcurPPlogg$ \\
~~~$a$ (AU)               \dotfill    & $\hatcurPParel$ \\
~~~$T_{\rm eq}$ (K)       \dotfill    & $\hatcurPPteff$ \\
~~~$\Theta$\tablenotemark{f}\dotfill  & $\hatcurPPtheta$ \\
~~~$\langle F \rangle$ ($10^{\hatcurPPfluxavgdim}$\ergscmsq) 
\tablenotemark{g}         \dotfill    & $\hatcurPPfluxavg$ \\
[-1.5ex]
\enddata
\tablenotetext{a}{
    \ensuremath{T_c}: Reference epoch of mid transit that minimizes the
    correlation with the orbital period. BJD is calculated from UTC.
    \ensuremath{T_{14}}: total transit duration, time between first to
    last contact;
    \ensuremath{T_{12}=T_{34}}: ingress/egress time, time between first
    and second, or third and fourth contact.
}
\tablenotetext{b}{
    Reciprocal of the half duration of the transit used as a jump
    parameter in our MCMC analysis in place of $\arstar$. It is
    related to $\arstar$ by the expression $\zrstar = \arstar
    (2\pi(1+e\sin \omega))/(P \sqrt{1 - b^{2}}\sqrt{1-e^{2}})$
    \citep{bakos:2010:hat11}.
}
\tablenotetext{c}{
        Values for a quadratic law given separately for the Sloan~$g$,
        $r$, and $i$ filters.  These values were adopted from the
        tabulations by \cite{claret:2004} according to the
        spectroscopic (SPC) parameters listed in \reftabl{stellar}.
}
\tablenotetext{d}{
    This jitter was added in quadrature to the RV uncertainties for
    each instrument such that $\chi^{2}/{\rm dof} = 1$ for the
    observations from that instrument. In the case of HDS, $\chi^{2}/{\rm dof} < 1$, so no jitter was added. 
}
\tablenotetext{e}{
    Correlation coefficient between the planetary mass \mpl\ and radius
    \rpl.
}
\tablenotetext{f}{
    The Safronov number is given by $\Theta = \frac{1}{2}(V_{\rm
    esc}/V_{\rm orb})^2 = (a/\rpl)(\mpl / \mstar )$
    \citep[see][]{hansen:2007}.
}
\tablenotetext{g}{
    Incoming flux per unit surface area, averaged over the orbit.
}
\end{deluxetable*}

\subsection{Constraining the radius--wavelength dependency of \hatcurb{}}
\label{sec:constr-radi-wavel}

Multi-band transit observations by GROND can provide constraints on the dependency between planet radius and wavelength \citep[e.g.][]{2013A&amp;A...553A..26N,2013MNRAS.436....2M}, and potentially probe for molecular absorption and Rayleigh scattering features in the transmission spectrum of a planet. We performed a separate fitting of the GROND full transit data from 2012 December 11, simultaneously fitting for the transit parameters $T_c$, $a/R_\star$, and $i$, and the individual $R_p/R_\star$ for each passband. The fitting is performed using the JKTEBOP eclipsing binary model \citep{Nelson1972,Southworth2004}, with both quadratic limb darkening coefficients fixed to that of \citet{claret:2004}, and freed and parameterised according to \citet{2013MNRAS.435.2152K}. The best fit parameters and uncertainties are explored by the \emph{emcee} implementation of a Markov chain Monte Carlo routine \citep{ForemanMackey2012} under Python. Simultaneous EPD is performed on the residuals for each iteration with a linear combination of the first order terms for time, target star X position, Y position, full width at half maximum, and airmass. The final $R_p/R_\star$ values are consistent with each other to within errors for the fixed and free limb darkening coefficient analyses. 

\begin{figure}
  \centering
  \includegraphics[width=9cm]{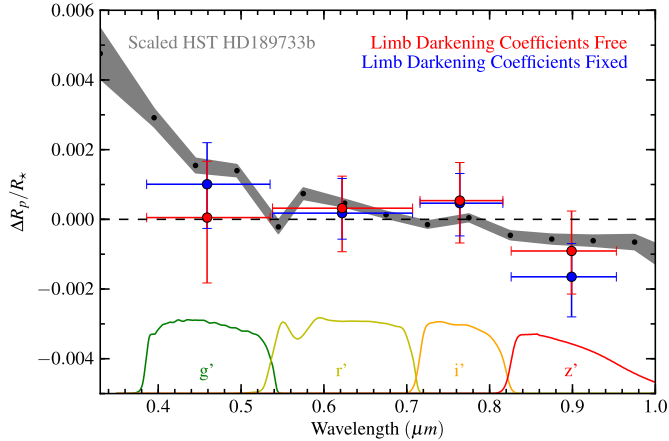}
  \caption{Variations in $R_p/R_\star$ over the optical passbands for the GROND full transit on 2012 December 11. A mean radius ratio has been subtracted for each passband. The radius ratios from the limb darkening fixed (blue) and free (red) analyses are plotted. We also plot the transmission spectrum of HD 189733b as observed using HST by \citet{2008MNRAS.385..109P,2011MNRAS.416.1443S}, and scaled to match the scale height and radius ratio of \hatcurb{}. The transmission curves for each filter are plotted at the bottom.}
  \label{fig:GROND_rprs}
\end{figure}

The deviation from mean radius for each passband is plotted in \reffigl{GROND_rprs}. For comparison, we also plot the wavelength-radius variation of HD 189733b, as measured using the Hubble Space Telescope (HST) by \citet{2008MNRAS.385..109P,2011MNRAS.416.1443S}, and scaled to match the scale height (500 km) and $R_p/R_\star$ of \hatcurb{}, assuming an H$_2$ dominated atmosphere \citep[following][]{2008A&amp;A...487..357S}. HD 189733b is a pL class planet according to \citet{2008ApJ...678.1419F}, with a mildy irradiated atmosphere that is potentially similar to that of \hatcurb. The results of the GROND observations are consistent with both a null detection of atmospheric features and that expected from the scaled measurements of HD 189733b. We do not see obvious star spot crossing events in the transit \lc, although unocculted spots can also cause a slope in the broadband $R_p/R_\star$ measurements \citep[e.g.][]{2008MNRAS.385..109P,2011MNRAS.416.1443S}. Whilst we do not detect any atmospheric features on \hatcurb, the large scale height makes \hatcurb{} an appealing target for future transmission spectroscopy observations. Future observations in the bluer $U$-band may also reveal opacity variations in the atmosphere by H$_2$ Rayleigh scattering \citep[e.g.][]{2011MNRAS.416.1443S,2013MNRAS.436.2956S,2013ApJ...778..184J,2013A&amp;A...559A..32N}.

\section{Discussion}
\label{sec:discussion}

\begin{figure}
  \centering
  \includegraphics[width=9cm]{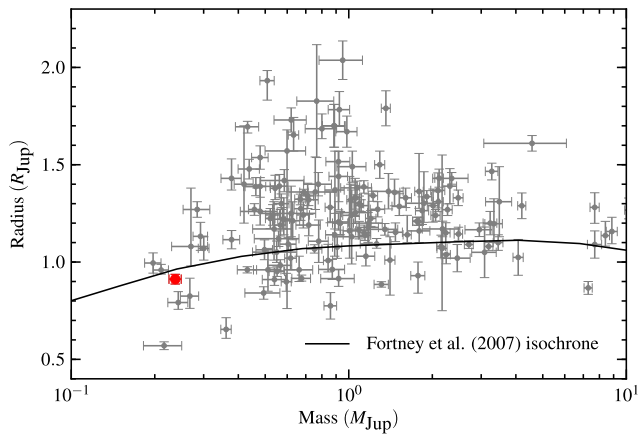}
  \caption{The mass--radius distribution of transiting gas giants\footnote{As of 12 Dec 2013, \url{exoplanets.org}} ($M_p > 0.1 \,\mjup$, $P<10$\,d) are plotted. \hatcurb{} is marked in red. Confirmed planets with masses and radii are plotted in gray. The isochrone from \citet{2007ApJ...659.1661F} for 4.5 Gyr old gas-giant planets, with $10\,M_\text{Earth}$ core sizes, orbiting 0.045 AU from the host star, is shown by solid line.}
  \label{fig:massradius}
\end{figure}

We presented the discovery of \hatcurb{}, a transiting hot-Saturn with mass of $\hatcurPPmlong\,\mjup$ and radius of $\hatcurPPrlong\,\rjup$. \hatcurb{} is the the lowest mass and radius planet to date reported by the HATSouth survey. The host star is a quiet, slowly rotating G-dwarf with a stellar mass of \hatcurISOmlong{}\,\msun{} and radius of \hatcurISOrlong{}\,\rsun. The mass and radius of \hatcurb{} are plotted in the context of existing close-in transiting gas giants in \reffigl{massradius}.

The radius of \hatcurb{} is consistent with the model of an irradiated gas-giant that formed via core accretion \citep{2007ApJ...659.1661F}. The radius is also consistent within $1\sigma$ to the empirical radius relationship for Saturn-mass planets from \citet{2012A&amp;A...540A..99E}.  We examine below the empirical factors that affect the radius of irradiated gas-giants.

\subsection{The $T_\text{eq}$--\feh--radius relationship}
\label{sec:teq-radi-depend}

A number of previous studies have investigated the relationship between the planet radius distribution, host star metallicity, and levels of insolation \citep[e.g.][]{2006A&amp;A...453L..21G,2011MNRAS.410.1631E,beky:2011,2012A&amp;A...540A..99E}. The factors that impact the radius of a gas giant should be mass dependent. For example, the level insolation should have a less significant impact on the radius of the denser, more massive gas giants and brown dwarfs than on the less dense Saturn-mass planets. Here, we revisit the mass dependence of the planet radius on the host star metallicity and the planet equilibrium temperature. 

\begin{figure}
  \centering
  \begin{tabular}{cc}
      \includegraphics[width=9cm]{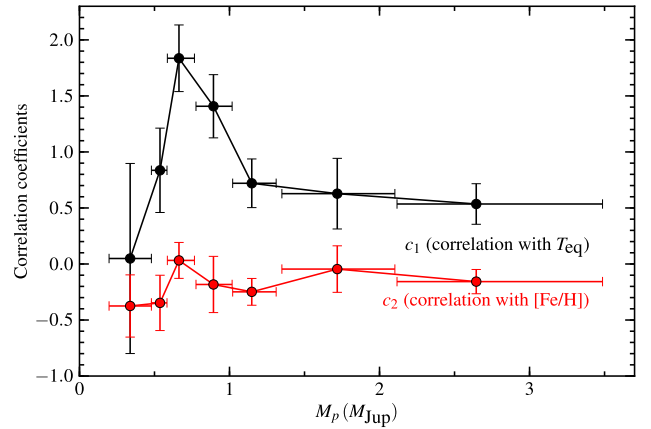}
  \end{tabular}

  \caption{The correlation between planet radius, equilibrium temperature and metallicity are plotted. For each mass bin of size 20, we calculate the correlation coefficients $c_1$ and $c_2$ (Equation~\ref{eq:dependence}). The vertical error bars are derived from bootstrapping the sample. The horizontal error bars show the extent of each mass bin. }
  \label{fig:radius-teq-feh}
\end{figure}

We bin the planet population into samples of 20, and perform a least squares fit for a linear dependence between radius, mass, equilibrium temperature $T_\text{eq}$, and metallicity:
\begin{align}
  \label{eq:dependence}
  R_\text{model} &= c_1 \mathcal{T} + c_2 \mathcal{M} + c_3 \log M_p + c_4\\
  \mathcal{T} &\equiv \frac{T_\text{eq}-T_\text{eq,mean}}{T_\text{eq,max}-T_\text{eq,min}} \notag \\
  \mathcal{M} &\equiv \frac{\feh-\feh_\text{mean}}{\feh_\text{max}-\feh_\text{min}}  \notag
\end{align}
where the magnitude of $c_1$ and $c_2$ are used to judge the level of correlation for $T_\text{eq}$ and [Fe/H], respectively. $c_3$ takes into account a linear dependence between mass and radius within the mass bin. $c_4$ is an arbitrary offset in the fit. The errors in the coefficients are derived by bootstrapping the analysis within each mass bin. Since each mass bin covers a relatively small mass range, a linear dependence is sufficient (see Figure~\ref{fig:radius-teq-feh} for the sizes of each mass bin). We find a peak in the mass dependence of the $T_\text{eq}$ correlation at $M_p \sim 1\,\mjup$, and a general lack of overall correlation between $R_p$ and \feh. The correlation coefficients are plotted against their respective mass bins in Figure~\ref{fig:radius-teq-feh}. We repeated the exercise using only planets with solar-mass hosts ($0.8 < M_\star < 1.2 \, \msun$), to reduce any potential selection effects in the target selection and spectral classifications of the surveys. Smaller planets are found in longer periods \citep[e.g.][]{2005MNRAS.356..955M,2009MNRAS.396.1012D}, biasing the $T_\text{eq}$-mass distribution. To reduce the effect of the bias, we re-perform the analysis using only mildy irradiated planets $(T_\text{eq} < 1500\,\text{K})$. In addition, the radius-$T_\text{eq}$ dependence is non-linear over the general population \citep{2011ApJS..197...12D}, limiting the $T_\text{eq}$ range has the added benefit of reducing effect of the non-linear dependence on the analysis. In all cases we find the peak dependence to $T_\text{eq}$ to be $\sim 1\,\mjup$, and a lack of dependence on [Fe/H]. 

We also perform the same analysis for the entire population of hot gas-giants, fitting for a second order polynomial in mass-radius, and linear dependence to $T_\text{eq}$ and \feh. We find a strong correlation in $T_\text{eq}$ with $c_1 = 0.81\pm0.17$, and an insignificant correlation in \feh{}, with $c_2 = -0.25\pm0.14$. Increasing the order of the polynomial does not affect the coefficient values within errors. We find the overall dependence to \feh{} to be weak at best. \citet{2011ApJ...736L..29M} suggests that the \feh-radius dependence is more prominent for the least irradiated planets, we limit the analysis to planets with $T_\text{eq} < 1500\,\text{K}$, but still find a lack of correlation with \feh, with $c_1 = 0.37\pm0.11$ and $c_2 = -0.15\pm 0.10$.

We find the radius of Saturn-mass planets are less affected by their equilibrium temperature than Jupiter-mass planets, in agreement with the Singular Value Decomposition analysis performed by \citet{2012A&amp;A...540A..99E}. In addition, we also find that the radius of planets with $M_p>1\,\mjup$ are also less dependent on equilibrium temperature. This effect is reproduced by the isochrones from \citet{2007ApJ...659.1661F}. The isochrones can also reproduce a drop in the correlation strength between irradiation and radius for the least massive gas-giants ($M_p<0.3\,\mjup$), but require the presence of a large core ($M_c > 10 \,M_\text{Earth}$). Interestingly, we find no statistically significant dependence of radius on the host star metallicity, contrary to previous examinations \citep[e.g.][]{2006A&amp;A...453L..21G,beky:2011,2011MNRAS.410.1631E,2012A&amp;A...540A..99E}. It is not clear how the host star metallicity affects the metallicity and radius of the planet. A higher metallicity disk may produce planets with more massive cores, leading to a smaller overall radius \citep[e.g.][]{2006A&amp;A...453L..21G}, but a higher opacity atmosphere is more efficient at retaining heat, reducing the rate of contraction, leading to a more inflated radius \citep[e.g.][]{2007ApJ...661..502B,2011ApJ...736...47B}.

\acknowledgements 
Development of the HATSouth project was funded by NSF MRI grant
NSF/AST-0723074, operations are supported by NASA grant NNX12AH91H, and
follow-up observations receive partial support from grant
NSF/AST-1108686.
Work at the Australian National University is supported by ARC Laureate
Fellowship Grant FL0992131.
Followup observations with the ESO~2.2\,m/FEROS instrument were
performed under MPI guaranteed time (P087.A-9014(A), P088.A-9008(A),
P089.A-9008(A)) and Chilean time (P087.C-0508(A)).
A.J. acknowledges support from
FONDECYT project 1130857, BASAL CATA PFB-06, and the Millennium
Science Initiative, Chilean Ministry of Economy (Millenium Institute
of Astrophysics MAS and Nucleus P10-022-F).
V.S.\ acknowledges support form BASAL CATA PFB-06. M.R. acknowledges support from FONDECYT postdoctoral fellowship No3120097. R.B.\ and N.E.\
acknowledge support from CONICYT-PCHA/Doctorado Nacional and Fondecyt
project 1130857.
This work is based on observations made with ESO Telescopes at the La
Silla Observatory under programme IDs P087.A-9014(A), P088.A-9008(A),
P089.A-9008(A), P087.C-0508(A), 089.A-9006(A), and
We acknowledge the use of the AAVSO Photometric All-Sky Survey (APASS),
funded by the Robert Martin Ayers Sciences Fund, and the SIMBAD
database, operated at CDS, Strasbourg, France.
Operations at the MPG/ESO 2.2\,m Telescope are jointly performed by the
Max Planck Gesellschaft and the European Southern Observatory.  The
imaging system GROND has been built by the high-energy group of MPE in
collaboration with the LSW Tautenburg and ESO\@. We thank Régis Lachaume for his technical assistance during the
observations at the MPG/ESO 2.2\,m Telescope.
Australian access to the Magellan Telescopes was supported through the
National Collaborative Research Infrastructure Strategy of the
Australian Federal Government.
We thank Albert Jahnke, Toni Hanke (HESS), Peter Conroy (MSO) for
their contributions to the HATSouth project.

\bibliographystyle{apj}
\bibliography{hatsbib}

\end{document}

%% file: newcommand_gen.tex
\newcommand{\ctbd}[1]{}

\newcommand{\lc}{light curve}
\newcommand{\lcs}{light curves}
\newcommand{\Lc}{Light curve}

\newcommand{\band}[1]{\ensuremath{#1}-band}

\newcommand{\kms}{\ensuremath{\rm km\,s^{-1}}}
\newcommand{\ms}{\ensuremath{\rm m\,s^{-1}}}

\newcommand{\gcmc}{\ensuremath{\rm g\,cm^{-3}}}
\newcommand{\ergscmsq}{\ensuremath{\rm erg\,s^{-1}\,cm^{-2}}}

\newcommand{\pxs}{\ensuremath{\rm \arcsec pixel^{-1}}}

\newcommand{\teff}{\ensuremath{T_{\rm eff}}}
\newcommand{\logg}{\ensuremath{\log{g}}}
\newcommand{\vsini}{\ensuremath{v \sin{i}}}
\newcommand{\feh}{\ensuremath{\rm [Fe/H]}}

\newcommand{\rsun}{\ensuremath{R_\sun}}
\newcommand{\msun}{\ensuremath{M_\sun}}
\newcommand{\lsun}{\ensuremath{L_\sun}}

\newcommand{\rstar}{\ensuremath{R_\star}}
\newcommand{\mstar}{\ensuremath{M_\star}}
\newcommand{\lstar}{\ensuremath{L_\star}}

\newcommand{\teffstar}{\ensuremath{T_{\rm eff\star}}}

\newcommand{\loggstar}{\ensuremath{\log{g_{\star}}}}

\newcommand{\rpl}{\ensuremath{R_{p}}}
\newcommand{\mpl}{\ensuremath{M_{p}}}

\newcommand{\rhopl}{\ensuremath{\rho_{p}}}

\newcommand{\arstar}{\ensuremath{a/\rstar}}
\newcommand{\zrstar}{\ensuremath{\zeta/\rstar}}

\newcommand{\rjup}{\ensuremath{R_{\rm J}}}
\newcommand{\mjup}{\ensuremath{M_{\rm J}}}

\newcommand{\reffigl}[1]{Figure~\ref{fig:#1}}
\newcommand{\refsecl}[1]{\mbox{Section \ref{sec:#1}}}

\newcommand{\reftabl}[1]{Table~\ref{tab:#1}}



%% file: newcommand_spec.tex
\input{HATS549-003.eccen.tex}
\newcommand{\hatcur}{HATS-5}
\newcommand{\hatcurb}{HATS-5b}

\newcommand{\hatcurSMEversion}{i}                                       
\newcommand{\hatcurSMEteff}{\ifthenelse{\equal{\hatcurSMEversion}{i}}{\hatcurSMEiteff}{\hatcurSMEiiteff}}
\newcommand{\hatcurSMEzfeh}{\ifthenelse{\equal{\hatcurSMEversion}{i}}{\hatcurSMEizfeh}{\hatcurSMEiizfeh}}
\newcommand{\hatcurSMEzfehshort}{\ifthenelse{\equal{\hatcurSMEversion}{i}}{\hatcurSMEizfehshort}{\hatcurSMEiizfehshort}}
\newcommand{\hatcurSMElogg}{\ifthenelse{\equal{\hatcurSMEversion}{i}}{\hatcurSMEilogg}{\hatcurSMEiilogg}}
\newcommand{\hatcurSMEvsin}{\ifthenelse{\equal{\hatcurSMEversion}{i}}{\hatcurSMEivsin}{\hatcurSMEiivsin}}
\newcommand{\hatcurSMEvmac}{\ifthenelse{\equal{\hatcurSMEversion}{i}}{\hatcurSMEivmac}{\hatcurSMEiivmac}}
\newcommand{\hatcurSMEvmic}{\ifthenelse{\equal{\hatcurSMEversion}{i}}{\hatcurSMEivmic}{\hatcurSMEiivmic}}

\newcommand{\hatcurSMEversioneccen}{i}                                       
\newcommand{\hatcurSMEteffeccen}{\ifthenelse{\equal{\hatcurSMEversioneccen}{i}}{\hatcurSMEiteffeccen}{\hatcurSMEiiteffeccen}}
\newcommand{\hatcurSMEzfeheccen}{\ifthenelse{\equal{\hatcurSMEversioneccen}{i}}{\hatcurSMEizfeheccen}{\hatcurSMEiizfeheccen}}
\newcommand{\hatcurSMEzfehshorteccen}{\ifthenelse{\equal{\hatcurSMEversioneccen}{i}}{\hatcurSMEizfehshorteccen}{\hatcurSMEiizfehshorteccen}}
\newcommand{\hatcurSMEloggeccen}{\ifthenelse{\equal{\hatcurSMEversioneccen}{i}}{\hatcurSMEiloggeccen}{\hatcurSMEiiloggeccen}}
\newcommand{\hatcurSMEvsineccen}{\ifthenelse{\equal{\hatcurSMEversioneccen}{i}}{\hatcurSMEivsineccen}{\hatcurSMEiivsineccen}}
\newcommand{\hatcurSMEvmaceccen}{\ifthenelse{\equal{\hatcurSMEversioneccen}{i}}{\hatcurSMEivmaceccen}{\hatcurSMEiivmaceccen}}
\newcommand{\hatcurSMEvmiceccen}{\ifthenelse{\equal{\hatcurSMEversioneccen}{i}}{\hatcurSMEivmiceccen}{\hatcurSMEiivmiceccen}}

%% file: HATS549-003.eccen.tex
\newcommand{\hatcurhtr}{HATS549-003}                                   
\newcommand{\hatcurCCra}{\ensuremath{04^{\mathrm h}28^{\mathrm m}53.47{\mathrm s}}}                                  
\newcommand{\hatcurCCdec}{\ensuremath{-21{\arcdeg}28{\arcmin}54.9{\arcsec}}}                                 
\newcommand{\hatcurCCgsc}{GSC~5897-00933}                              
\newcommand{\hatcurCCapassmVshort}{12.6}                               
\newcommand{\hatcurCCapassmV}{\ensuremath{12.630\pm0.030}}             
\newcommand{\hatcurCCapassmB}{\ensuremath{13.439\pm0.010}}             
\newcommand{\hatcurCCtwomassJmag}{\ensuremath{11.199\pm0.023}}         
\newcommand{\hatcurCCtwomassHmag}{\ensuremath{10.772\pm0.023}}         
\newcommand{\hatcurCCtwomassKmag}{\ensuremath{10.703\pm0.023}}         
\newcommand{\hatcurLCrprstar}{\ensuremath{0.1076\pm0.0004}}            
\newcommand{\hatcurLCimp}{\ensuremath{0.158_{-0.064}^{+0.057}}}        
\newcommand{\hatcurLCzeta}{\ensuremath{17.85\pm0.03}}                  
\newcommand{\hatcurLCdur}{\ensuremath{0.1244\pm0.0004}}                
\newcommand{\hatcurLCingdur}{\ensuremath{0.0124\pm0.0003}}             
\newcommand{\hatcurLCP}{\ensuremath{4.763387\pm0.000010}}              
\newcommand{\hatcurLCPprec}{\ensuremath{4.7633872}}                    
\newcommand{\hatcurLCPshort}{\ensuremath{4.7634}}                      
\newcommand{\hatcurLCT}{\ensuremath{2456273.79068\pm0.00010}}          
\newcommand{\hatcurSMEiteff}{\ensuremath{5304\pm50}}                   
\newcommand{\hatcurSMEizfeh}{\ensuremath{0.19\pm0.08}}                 
\newcommand{\hatcurSMEizfehshort}{\ensuremath{0.19}}                   
\newcommand{\hatcurSMEilogg}{\ensuremath{4.51\pm0.10}}                 
\newcommand{\hatcurSMEivsin}{\ensuremath{0.8\pm0.5}}                   
\newcommand{\hatcurSMEivmac}{\ensuremath{NULL}}                        
\newcommand{\hatcurSMEivmic}{\ensuremath{NULL}}                        
\newcommand{\hatcurSMEiiteff}{\ensuremath{NULL\pmNULL}}                
\newcommand{\hatcurSMEiizfeh}{\ensuremath{NULL\pmNULL}}                
\newcommand{\hatcurSMEiizfehshort}{\ensuremath{NULL}}                  
\newcommand{\hatcurSMEiilogg}{\ensuremath{NULL\pmNULL}}                
\newcommand{\hatcurSMEiivsin}{\ensuremath{NULL\pmNULL}}                
\newcommand{\hatcurSMEiivmac}{\ensuremath{NULL}}                       
\newcommand{\hatcurSMEiivmic}{\ensuremath{NULL}}                       
\newcommand{\hatcurLBii}{\ensuremath{0.3571}}                          
\newcommand{\hatcurLBiii}{\ensuremath{0.2823}}                         
\newcommand{\hatcurLBir}{\ensuremath{0.4723}}                          
\newcommand{\hatcurLBiir}{\ensuremath{0.2542}}                         
\newcommand{\hatcurLBiR}{\ensuremath{0.4405}}                          
\newcommand{\hatcurLBiiR}{\ensuremath{0.2628}}                         
\newcommand{\hatcurISOm}{\ensuremath{0.94\pm0.03}}                     
\newcommand{\hatcurISOmshort}{\ensuremath{0.94}}                       
\newcommand{\hatcurISOmlong}{\ensuremath{0.936\pm0.028}}               
\newcommand{\hatcurISOr}{\ensuremath{0.87\pm0.02}}                     
\newcommand{\hatcurISOrshort}{\ensuremath{0.87}}                       
\newcommand{\hatcurISOrlong}{\ensuremath{0.871\pm0.023}}               
\newcommand{\hatcurISOlogg}{\ensuremath{4.53\pm0.02}}                  
\newcommand{\hatcurISOlum}{\ensuremath{0.54\pm0.04}}                   
\newcommand{\hatcurISOmv}{\ensuremath{5.60\pm0.09}}                    
\newcommand{\hatcurISOage}{\ensuremath{3.6_{-1.9}^{+2.6}}}             
\newcommand{\hatcurISOMK}{\ensuremath{3.69\pm0.06}}                    
\newcommand{\hatcurISOspec}{G}                                         
\newcommand{\hatcurRVK}{\ensuremath{30.0\pm1.4}}                       
\newcommand{\hatcurRVrk}{\ensuremath{0.017\pm0.090}}                   
\newcommand{\hatcurRVrh}{\ensuremath{-0.015\pm0.121}}                  
\newcommand{\hatcurRVk}{\ensuremath{0.001\pm0.016}}                    
\newcommand{\hatcurRVh}{\ensuremath{-0.001\pm0.024}}                   
\newcommand{\hatcurRVjitterA}{\ensuremath{2.0}}                        
\newcommand{\hatcurRVjitterB}{\ensuremath{0.0}}                        
\newcommand{\hatcurRVeccen}{\ensuremath{0.019\pm0.019}}                
\newcommand{\hatcurRVomega}{\ensuremath{204\pm107}}                    
\newcommand{\hatcurPPi}{\ensuremath{89.3\pm0.3}}                       
\newcommand{\hatcurPPg}{\ensuremath{7.1\pm0.5}}                        
\newcommand{\hatcurPPlogg}{\ensuremath{2.85\pm0.03}}                   
\newcommand{\hatcurPPar}{\ensuremath{13.38\pm0.34}}                    
\newcommand{\hatcurPParel}{\ensuremath{0.0542\pm0.0006}}               
\newcommand{\hatcurPPrho}{\ensuremath{0.39\pm0.04}}                    
\newcommand{\hatcurPPm}{\ensuremath{0.24\pm0.01}}                      
\newcommand{\hatcurPPmshort}{\ensuremath{0.24}}                        
\newcommand{\hatcurPPmlong}{\ensuremath{0.237\pm0.012}}                
\newcommand{\hatcurPPr}{\ensuremath{0.91\pm0.03}}                      
\newcommand{\hatcurPPrshort}{\ensuremath{0.91}}                        
\newcommand{\hatcurPPrlong}{\ensuremath{0.912\pm0.025}}                
\newcommand{\hatcurPPmrcorr}{\ensuremath{-0.01}}                       
\newcommand{\hatcurPPteff}{\ensuremath{1025\pm17}}                     
\newcommand{\hatcurPPtheta}{\ensuremath{0.030\pm0.002}}                
\newcommand{\hatcurPPfluxavg}{\ensuremath{2.50\pm0.17}}                
\newcommand{\hatcurPPfluxavgdim}{\ensuremath{8}}                       
\newcommand{\hatcurXdist}{\ensuremath{257\pm8}}                        

%% file: data/rvtable.tex
$ 56190.07247 $ & \nodata      & \nodata      & $  -12.2 $ & $   32.1 $ & $   0.425 $ & Subaru (I$_2$ free) \tablenotemark{c} \\
$ 56190.08373 $ & \nodata      & \nodata      & $  -15.3 $ & $   29.4 $ & $   0.427 $ & Subaru (I$_2$ free)\\
$ 56190.09502 $ & \nodata      & \nodata      & $  -19.4 $ & $   31.8 $ & $   0.429 $ & Subaru (I$_2$ free)\\
$ 56191.05697 $ & $    24.85 $ & $     6.66 $ & $   12.1 $ & $   29.7 $ & $   0.631 $ & Subaru \\
$ 56191.07042 $ & $    18.93 $ & $     7.04 $ & $   18.7 $ & $   30.8 $ & $   0.634 $ & Subaru \\
$ 56191.08294 $ & $    24.24 $ & $     5.97 $ & $  -14.8 $ & $   31.7 $ & $   0.637 $ & Subaru \\
$ 56192.05128 $ & $    23.95 $ & $     5.36 $ & $   10.3 $ & $   30.5 $ & $   0.840 $ & Subaru \\
$ 56192.06253 $ & $    25.71 $ & $     5.57 $ & $   -0.7 $ & $   30.2 $ & $   0.842 $ & Subaru \\
$ 56192.07379 $ & $    18.94 $ & $     6.40 $ & $   18.5 $ & $   25.1 $ & $   0.845 $ & Subaru \\
$ 56193.04260 $ & $     2.96 $ & $     6.68 $ & $  -20.6 $ & $   32.5 $ & $   0.048 $ & Subaru \\
$ 56193.05386 $ & $    -6.50 $ & $     6.59 $ & $   -1.5 $ & $   26.4 $ & $   0.051 $ & Subaru \\
$ 56193.06522 $ & $   -11.85 $ & $     5.82 $ & $   -4.7 $ & $   31.1 $ & $   0.053 $ & Subaru \\
$ 56289.54439 $ & $   -31.21 $ & $     3.04 $ & \nodata      & \nodata      & $   0.307 $ & PFS \\
$ 56289.55987 $ & $   -27.51 $ & $     3.19 $ & \nodata      & \nodata      & $   0.311 $ & PFS \\
$ 56290.68128 $ & $     5.01 $ & $     2.26 $ & \nodata      & \nodata      & $   0.546 $ & PFS \\
$ 56290.69567 $ & $    11.78 $ & $     2.39 $ & \nodata      & \nodata      & $   0.549 $ & PFS \\
$ 56291.70608 $ & $    26.14 $ & $     2.45 $ & \nodata      & \nodata      & $   0.761 $ & PFS \\
$ 56291.72039 $ & $    35.20 $ & $     2.67 $ & \nodata      & \nodata      & $   0.764 $ & PFS \\
$ 56292.65654 $ & $     5.09 $ & $     2.78 $ & \nodata      & \nodata      & $   0.961 $ & PFS \\
$ 56292.67135 $ & $    12.18 $ & $     2.99 $ & \nodata      & \nodata      & $   0.964 $ & PFS \\
$ 56344.52893 $ & $    20.99 $ & $     2.92 $ & \nodata      & \nodata      & $   0.850 $ & PFS \\
$ 56344.54311 $ & $    26.12 $ & $     2.91 $ & \nodata      & \nodata      & $   0.853 $ & PFS \\
$ 56355.51803 $ & $   -28.79 $ & $     3.80 $ & \nodata      & \nodata      & $   0.157 $ & PFS \\
$ 56355.53239 $ & $   -19.60 $ & $     4.73 $ & \nodata      & \nodata      & $   0.160 $ & PFS \\

%% file: data/phfu_tab_short.tex
$ 55547.37447 $ & $  -0.01041 $ & $   0.00339 $ & $ \cdots $ & $ r$ &         HS\\
$ 55456.87046 $ & $   0.00039 $ & $   0.00321 $ & $ \cdots $ & $ r$ &         HS\\
$ 55485.45085 $ & $  -0.00200 $ & $   0.00337 $ & $ \cdots $ & $ r$ &         HS\\
$ 55518.79459 $ & $  -0.00569 $ & $   0.00345 $ & $ \cdots $ & $ r$ &         HS\\
$ 55499.74154 $ & $  -0.00095 $ & $   0.00326 $ & $ \cdots $ & $ r$ &         HS\\
$ 55461.63555 $ & $  -0.00360 $ & $   0.00332 $ & $ \cdots $ & $ r$ &         HS\\
$ 55547.37789 $ & $   0.00305 $ & $   0.00349 $ & $ \cdots $ & $ r$ &         HS\\
$ 55456.87378 $ & $   0.00536 $ & $   0.00324 $ & $ \cdots $ & $ r$ &         HS\\
$ 55518.79794 $ & $  -0.00583 $ & $   0.00346 $ & $ \cdots $ & $ r$ &         HS\\
$ 55499.74522 $ & $   0.00102 $ & $   0.00328 $ & $ \cdots $ & $ r$ &         HS\\